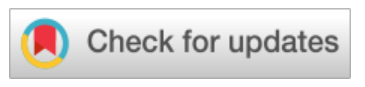

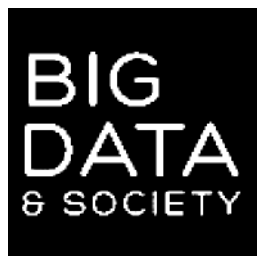

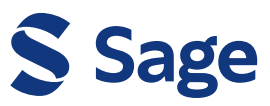

# From human-centered to social-centered artificial intelligence: Assessing ChatGPT's impact through disruptive events

**Skyler Wang[1], Ned Cooper[2] and Margaret Eby[3]**

### Abstract

Large language models (LLMs) and dialogue agents represent a significant shift in artificial intelligence (AI) research, particularly with the recent release of the GPT family of models. ChatGPT's generative capabilities and versatility across technical and creative domains led to its widespread adoption, marking a departure from more limited deployments of previous AI systems. While society grapples with the emerging cultural impacts of this new societal-scale technology, critiques of ChatGPT's impact within machine learning research communities have coalesced around its performance or other conventional safety evaluations relating to bias, toxicity, and "hallucination." We argue that these critiques draw heavily on a particular conceptualization of the "human-centered" framework, which tends to cast atomized individuals as the key recipients of technology's benefits and detriments. In this article, we direct attention to another dimension of LLMs and dialogue agents' impact: their effects on *social groups, institutions, and accompanying norms and practices*. By analyzing ChatGPT's social impact through a *social-centered* framework, we challenge individualistic approaches in AI development and contribute to ongoing debates around the ethical and responsible deployment of AI systems. We hope this effort will call attention to more comprehensive and longitudinal evaluation tools (e.g., including more ethnographic analyses and participatory approaches) and compel technologists to complement human-centered thinking with social-centered approaches.

## Introduction

ChatGPT, a generative dialogue agent created by OpenAI, launched in November 2022. By January 2023, it had reached 100 million monthly active users, quickly becoming "the fastest-growing consumer application in history" at that time (Hu, 2023). Since then, journalists and social scientists have written extensively about ChatGPT's widespread adoption and the social impact it has engendered. Schools and teachers rushed to adapt their classroom policies and practices to prevent students from using the dialogue agent to cheat on their assignments (Castillo, 2023; Roose, 2023). As employees increasingly relied on ChatGPT to perform critical business tasks (Nolan, 2023), many corporations and organizations scrambled to rewrite their workplace policies to ensure that sensitive, confidential, or proprietary data were not transferred to OpenAI through the system's web interface. The rapid reach and influence of ChatGPT are unparalleled—thousands of downstream applications incorporating OpenAI's API (application programming interface) and an expansive spawning of similarly competitive large language models (LLMs) and dialogue agents followed its inception.

While society grapples with the emerging cultural impacts of ChatGPT, machine learning research communities are focused on a different set of metrics. Invocations of "impact" in the documentation accompanying the release of models tend to focus on the risk of harm inflicted on *individual users* from a performance and safety standpoint (e.g., Borji, 2023). Generally, high-performance LLM

[1]Department of Sociology, McGill University, Montreal, Quebec, Canada
[2]School of Cybernetics, Australian National University, Canberra, ACT, Australia
[3]Department of Sociology, University of California, Berkeley, CA, USA

The first two authors contributed equally.

**Corresponding author:**
Ned Cooper, School of Cybernetics, Australian National University, 35 Science Road, Acton 2601, Australia.
Email: edward.cooper@anu.edu.au

outputs reflect user intentions, are factually accurate, and minimize "hallucination"[1] (i.e., the tendency to generate inaccurate outputs unfaithful to the training data; Bai et al., 2022; Ouyang et al., 2022; Thoppilan et al., 2022). Meanwhile, safety evaluations are broken down into two central tenets: 1) assessing the system for fairness, bias, and toxicity, and 2) ensuring sufficient system guardrails to prevent misuse (Bai et al., 2022; Ouyang et al., 2022; Thoppilan et al., 2022). We argue that the focus on individual users and their psyches, which draws on a particular, individualistic view of *human-centered artificial intelligence* (AI), is a double-edged sword; while it prioritizes users' capacities and needs during system development and evaluation, human–machine alignment under such a configuration atomizes and dislodges users from their social contexts (Selbst et al., 2019).

The propensity for individual-level assessments accompanying the release of models assumes that taking care of the individual takes care of society and obscures the urgent need to examine and respond to second-order impacts (Joyce et al., 2021). To address this critical gap, we argue that those deploying and evaluating *societal-scale* technologies such as ChatGPT must consider their effects on *social groups, institutions, and their norms and practices*. Using three "disruptive events" engendered by ChatGPT as points of discussion, we illustrate how current evaluation and assessment methods fail to consider the distribution of benefits of AI systems situated in group and institutional contexts. Building on these insights, we then propose concrete and actionable suggestions for machine learning researchers and practitioners to go beyond individual-level thinking in current deployments of human-centeredness and prioritize methods and techniques that take AI into a social-centered era. We hope the methods and approaches to conducting social impact assessments discussed in this article can guide and alter how machine learning researchers and practitioners work on a fundamental level, offering a framework for realizing social-centered AI in practice.

## The multiple levels of AI impact

Below, we first introduce the emergence of "human-centeredness" and spotlight the predominant interpretation of the concept in existing machine learning workflows. Then, we introduce the concept of "social-centeredness" to complement and transcend this dominant view of human-centeredness.

### *Human-centeredness: impact on individuals*

Human-centered technology arrived in the 1980s as a response to the foundering of technology-centered thinking. As technologists sought to prioritize the "human factor" (Vicente, 2010), the move toward human-centeredness positioned human operators as an asset and saw human–computer interaction as complementary rather than adversarial. In the 1990s, the emergence of value-sensitive design (Friedman, 1996) expanded human-centeredness to consider the values and behaviors of indirect stakeholders, not only direct users of technology.

As a guiding principle, human-centeredness gives machine learning researchers and practitioners an accessible vocabulary to articulate how systems are responsibly built. The term "human-centered" is found not only in the publications of LLMs but also across the websites of prominent organizations such as Google (Croak, 2023), IBM (Geyer, 2022), and AI4GOOD (Lamoutte, 2022). Furthermore, as evidenced by the volume of papers and workshops, interest in human-centered AI research has multiplied within prominent machine learning and computing conferences such as NeurIPS (Conference and Workshop on Neural Information Processing Systems) and CHI (Conference on Human Factors in Computing Systems) in recent years.

Human-centered AI is an umbrella term for a wide research agenda (Shneiderman, 2020). Definitionally, Riedl (2019: 33) proposes the following: "a perspective on AI and ML that algorithms must be designed with awareness that they are part of a larger system consisting of humans." Xu (2019: 42), on the other hand, decomposes the concept into three main components: 1) ethically aligned design, 2) technology that fully reflects human intelligence, and 3) human factors design "to ensure that AI solutions are explainable, comprehensible, useful, and usable." Proposing another triptych, Landay (2023) argues that true human-centered AI development must be "user-centered, community-centered, and societally-centered." In a recent review of research under the umbrella of human-centered AI, Capel and Brereton (2023) found an incredible breadth of work across four clusters: "Explainable and Interpretable AI," "Approaches to Design and Evaluate AI," "Human Teaming with AI," and "Ethical AI." These broad and varied conceptualizations suggest that interpretations of human-centered AI may largely be contingent on researchers' training, interest, and embedded networks.

Despite the variance, one area where human-centered AI has found relative consistency is in its evaluation methods. In the context of LLMs and dialogue agents, invocations of human-centeredness have coalesced around the sourcing of individual user feedback. In *Human-Centered AI*, Shneiderman (2022: 9) argues that what sets "regular" AI apart from human-centered AI is that the latter relies on an array of "user experience design methods," such as "user observation" and "usability testing," to build products that "amplify, augment, empower, and enhance human performance." These methods, which focus on soliciting individual feedback, have become standard practice in AI development, and they often rely on the assumption that the utterances and behaviors of a user are generalizable to the larger collectives to which they belong. Such a tendency, we argue, reflects a particular view of human-centered AI—that the broader pursuit of fairness and accountability rests on understanding individualistic behaviors.

## How evaluations of ChatGPT reflect individualistic human-centered AI

The way GPT models were evaluated for their impact—combining reinforcement learning from human feedback and expert red-teaming—exemplifies the individualistic conceptualization of the human-centered approach to AI development. This approach to LLM alignment reverberates across the industry (Bai et al., 2022; Glaese et al., 2022; Liang et al., 2022) and is rooted in machine learning's affinity with fields such as cognitive science, psychology, and neuroscience, all of which skew toward the individual as their primary unit of analysis (Hutchins, 1995). Under the current configuration, users, atomized, and removed from their larger social contexts (Selbst et al., 2019) are farmed for their feedback—an instrumental resource used for iterative development processes that give machine learning researchers and practitioners the ability to control how their systems affect people.

The initial version of ChatGPT ran on GPT3.5, which was fine-tuned for conversational dialogue. GPT3.5, the base model of the free version of ChatGPT, was trained with human feedback to align model outputs with user intent (OpenAI, 2022). More specifically, GPT3.5 was initially evaluated against three criteria: helpfulness (following user instructions), truthfulness (the tendency to "hallucinate"), and harmlessness (toxicity and bias of outputs; Ouyang et al., 2022). To build a tool for detecting harmful content produced by ChatGPT, OpenAI relied in part on labeled examples of toxic language produced by outsourced laborers based in Kenya (Perrigo, 2023).

Following initial training and evaluation, OpenAI focused on "red teaming" ChatGPT before public release—a form of evaluation in which users prompt a dialogue agent to elicit undesirable model behaviors. OpenAI staff implemented red teaming in conjunction with external collaborators, though it is unclear how ChatGPT was updated based on their feedback prior to the initial release (Heaven, 2023). The company then continued to use expert red teaming before the release of GPT4 (Sanderson, 2023), recruiting more than 50 external experts to "qualitatively probe, adversarially test, and generally provide feedback on the GPT4 models" over a period of six months before its release (OpenAI, 2023a: 4). Despite the interdisciplinary nature of this expert congregation, comprising those trained in "alignment research, industry trust and safety, dis/misinformation, chemistry, biorisk, cybersecurity, nuclear risks, economics, human–computer interaction, law, education, and healthcare" (OpenAI, 2023a: 5), considerations of societal impact in GPT4's System Card pale in comparison to the documentation of engineering-centric mitigation efforts directed at improving performance for individual users.

ChatGPT's public release was also part of OpenAI's approach to align the system with human preferences "through an iterative process where [OpenAI] deploy, get feedback, and refine" (Heaven, 2023). Accordingly, the web interface included a mechanism for users to provide feedback to OpenAI on the outputs of the model—by rating an answer to a prompt with a thumb up or down, indicating whether a regenerated response is "better," the "same," or "worse" than the original, and giving additional open-ended feedback. OpenAI used feedback on nonfactual responses acquired through this mechanism, with other labeled comparison data, to train reward models in order to reduce GPT4's tendency to "hallucinate" relative to GPT3.5 (OpenAI, 2023a: 4).

As ChatGPT and downstream apps supported by its API are used to perform a wide range of tasks across social contexts, we need to expand our conceptualization of and evaluation strategies around "impact" to include a more *social* dimension. As Dourish (2006) reminds us, a superficial understanding of grounded scenarios often leads to technology design that fails to meet real-world deployment needs. Adapting to this orientation requires a shift away from interrogating individualized interactions between user and machine, and toward anticipating and understanding the effects of those interactions on human relations in complex and situated environments (Green and Viljoen, 2020; Shelby et al., 2023).

## Social-centeredness: impact on groups and institutions

In this section, we differentiate impacts at the societal level from those at the level of individual users—two dimensions varying in scale. More specifically, scale is not only a measure of the size of training datasets or the number of parameters of an LLM but scale is also a measure of the infrastructure required to deploy the "same" system and user experience across political, economic, geographical and cultural boundaries (Tsing, 2012; Young et al., 2024). In contrast to AI systems deployed for specific purposes and use cases, societal-scale systems such as ChatGPT impact social groups and institutions across such boundaries (Cooper, 2023). While all algorithmic systems influence and are influenced by their users and the social groups and institutions in which those users are embedded (Matias, 2023), some impacts may only emerge with such large-scale deployment (Weidinger et al., 2023).

For definitional clarity, we define groups as collectives of persons "characterized by shared place, common identity, collective culture, or social relations" (Fine, 2012: 160). Research shows that individuals are affected by group norms or scripts and that people behave differently when embedded within different groups (Feldman, 1984; Postmes et al., 2000). In other words, groups not only give individuals a sense of belonging but also help individuals decide different courses of action in the various domains of their lives.

Conversely, institutions are more abstract entities encompassing "systems of established and prevalent social rules that structure social interactions" (Hodgson, 2006: 2). The family, the state, religion, and law are all examples of institutions. While they emerge from "the thoughts and activities of individuals," they are "not reducible to them" (Hodgson, 2006: 2). Crucially, institutions are durable—they stabilize expectations and interactions between people, which in turn undergird convention and culture (Haveman and Wetts, 2019). That said, institutions may undergo shifts during unsettled times (Swidler, 1986), where social norms are in flux due to the arrival of new sociopolitical ideas or technologies. ChatGPT's arrival marks such a moment.

That groups and institutions originate from social interactions underscores the need to evaluate both the effects of user-system interactions *and* the impacts of such systems on social interactions outside the immediate user-system relationship. In other words, how does the use of ChatGPT change the norms and practices of the law, higher education, or the workplace? So far, individual-centric methods favored by machine learning research communities have largely ignored this question.

Incorporating groups and institutions into social impact considerations does not, however, mean we take a "net benefits" approach and apply it to new units of analysis. Exemplified by the question "Is ChatGPT Good or Bad for Society?" (Kim, 2023), the "net benefit" of an intervention or system is determined by weighing positive and negative outcomes against one another. When considering societal-scale technologies such as ChatGPT, weighing positive and negative outcomes for these entities is not only logistically challenging but also risks generalizing disparate impacts, overlooking how different groups and institutions may experience vastly different or contradictory consequences.

We seek to disrupt the dominant conceptualization of human-centered AI in recent deployments of LLMs and dialogue agents by steering attention toward a social-centered and equity-based framework, which examines the distribution of benefits in situated group and institutional contexts. Prioritizing groups and institutions compels us to confront how people, with their distinctive know-how and resources, use technologies to undertake work and navigate their social lives. For instance, in the college setting, what kinds of students are more equipped to use ChatGPT for their homework assignments, and who is more likely to be caught for plagiarism? How might we better understand instructors' responses, with varying levels of technical know-how, in their attempts to adapt their pedagogical approaches to meet the demands of new learning environments? If machine learning research communities continue to primarily rely on analyzing micro level effects between users and dialogue agents, we will remain ill-equipped to address these questions.

## The emerging social impacts of ChatGPT

Well-established processes exist for conducting anticipatory and concurrent social impact assessments in other disciplines. For example, health (World Health Organization, 2023), environmental (Glasson and Therivel, 2013), and human rights impact assessments (Kemp and Vanclay, 2013) all mandate a process for identifying, predicting, monitoring, and responding to the impacts of policies, programs, or projects on a population or environment. While impact assessments draw primarily on qualitative methods to monitor to what extent and how an intervention addresses the complex issues it is intended to confront, such assessments also often include quantitative techniques and participatory approaches, and may be conducted before undertaking an intervention.

Despite ongoing development, social impact assessments are relatively new to machine learning research communities. Recent toolkits (Krafft et al., 2021), frameworks (Raji et al., 2020), and studies (Costanza-Chock et al., 2022) of algorithmic auditing advocate for the involvement of affected people and communities early and throughout auditing processes. However, as these groups are typically indirect stakeholders rather than direct users, there are challenges in engaging them through traditional user-centric audit and evaluation methods. While methodological gaps remain, the social-centered approach offers an alternative to current practices of human-centered AI, and some of the major differences between these two paradigms are outlined in Table 1.

With these considerations in mind, we take the recent deployment of ChatGPT as our point of departure, showcasing three disruptive events to illustrate how ChatGPT's widespread adoption has impacted three societal domains: law, education, and work. We draw on Aquino et al.'s (2022: 1) definition of disruptive events as those that have "significant consequences for [those] who experience them, but [their] effects do not occur equally across the population." Importantly, these events do not just affect individuals at a micro level—they implicate different groups and institutions, alongside their norms and practices, in disparate ways (Aquino et al., 2022). In the context of algorithmic systems, we describe a disruptive event as an instance in which regular, institutionalized means of navigating a challenge are no longer possible due to the introduction of a new system.

These illustrative case studies, drawn from disruptive events widely discussed in popular media between March

**Table 1.** Key differences between human-centered and social-centered AI.

| | Human-centered AI* | Social-centered AI |
|---|---|---|
| Primary level of analysis | Micro (individual) | Meso (organizational) and macro (societal) |
| Population | Direct users | Direct and indirect users |
| Relational focus | Human–AI interaction | Human–human (e.g., between colleagues or employees and employers) or human-institution interactions (e.g., how humans interact with the law, education, etc.) |
| Main tools | User-centric evaluation and mitigation techniques (to remove bias, toxicity, or hallucination) | Ethnography, participatory approaches, and large-scale quantitative analyses |
| Impact analysis time frame | Mostly short-term | Both short- and long-term |

*We recognize that the definitions of human-centered AI are polysemic. This table's entries reflect the view of human-centered AI described in the previous section.

and May 2023, highlight the discrepancy between the matters considered in technical evaluations of ChatGPT and the concerns of the broader public regarding changes to social norms and practices. The cases received broad media engagement not because they were stories of technology but more so due to their human and social implications.

In our analysis of each case, we reflect on how ChatGPT disrupts norms and practices in each societal domain. The cases are broad in scope and continue to evolve, though they provide early indications of disruptions that machine learning research communities and social scientists alike ought to pay attention to in future and ongoing social impact assessments. Neither the domain nor the analysis of the societal impacts in each case is exhaustive, but we use these case studies to demonstrate how one might apply social-centered thinking when evaluating AI impact. While we cannot entirely eliminate unintended consequences, we can try to understand when and how technologies might prompt changes in behaviors and values within groups and institutions (Selbst et al., 2019). In each case, we emphasize how the relative access to resources among users or indirect stakeholders within and across affected groups or institutions is key for determining the uneven distribution of benefits and harms provided by LLMs and dialogue agents. At its core, social-centered AI is about mitigating inequalities in developing and deploying AI systems.

Finally, we recognize that systems such as ChatGPT can bring about several kinds of impacts on different entities, but our critical approach underscores effects of the negative kind. Positive impacts are more likely to be anticipated and devised by machine learning researchers and practitioners, while negative impacts—more costly and disruptive—tend to be ignored or downplayed (Ashurst et al., 2022; Liu et al., 2022). We also acknowledge that the "societal impact analysis" presented below is limited to qualitative media analysis and our interpretation of events; accordingly, we include a positionality statement at the end of the paper to add context to our standpoints.

## *Artificial intelligence chatbot's first defamation lawsuit: impact on law*

*Case study.* In November 2022, Brian Hood, an elected Mayor from Hepburn Shire Council in Victoria, Australia, received news from concerned voters that ChatGPT claimed he was involved in a foreign bribery scandal in the early 2000s involving a banknote printing business called "Securency," a subsidiary of the Reserve Bank of Australia (Reuters, 2023). When asked, "What role did Brian Hood have in the Securency bribery saga," ChatGPT erroneously "hallucinated" details, claiming that Hood was "charged with three counts of conspiracy to bribe foreign officials in Indonesia and Malaysia" and "one count of false accounting," pleaded guilty in 2012, and was "sentenced to two years and three months in prison" (Bonyhady, 2023).

Except none of it was true (Sands, 2023). In fact, Hood, an ex-employee of the subsidiary, was the whistleblower responsible for notifying authorities and exposing the international scandal in the first place. Hood was "shocked" and "angry" when he learned about this misinformation. His lawyers sent a "concerns notice" to OpenAI on March 21, 2023, "the first formal step to commencing defamation proceedings" (Bonyhady, 2023).

Hood's case was the first time someone in Australia had lodged a defamation suit against ChatGPT or, more generally, AI (Bonyhady, 2023). Ushering this case to trial would have meant the judicial system had to decide whether the creators of a dialogue agent could be held legally responsible if the agent produced defamatory statements about an individual. To win the case, Hood would have to prove that enough people had seen the prompt completion to constitute "serious harm." Hood's lawyers sent a letter to OpenAI demanding that ChatGPT cease sharing false information about him, and while the company rejected his defamation claim, any prompt about him now results in an error message from ChatGPT (Linebaugh, 2024). But as one recent report found, asking about the scandal itself may generate false information about Hood's involvement—OpenAI's fix has done little to address the root issues underlying potential defamation (Linebaugh, 2024).

### Societal impact analysis

When accessing ChatGPT's web interface, users are reminded that the system "may produce inaccurate information about people, places, or facts." When there are gaps in its training data, ChatGPT "hallucinates" to maintain conversational flow; as a result, it sometimes gets critical details wrong. Because ChatGPT's outputs do not come with confidence scores, those with less experience with dialogue agents or lower levels of digital literacy may take false information at face value. "Hallucination" is not unique to ChatGPT—many LLMs that emerged prior can generate falsehoods that people can spread and weaponize. However, the massive, global adoption of ChatGPT in a short period inscribed it with a form of legitimacy few AI systems have enjoyed. This status, incidentally, also invites greater legal scrutiny.

Hood's case compels us to ask: what happens when ChatGPT's fabricated outputs—or misinformation—begin making inroads in society to impact different people's lives? And how might the way such trends intersect with the law offer a look into how institutional disruptions within the legal and judicial systems are underway?

Although defamation laws vary across jurisdictions (Johnson, 2017), defamation generally refers to false written or verbal statements that could damage a third party's reputation. When applying defamation laws, legal entities try to balance safeguarding free speech with protecting individual reputations from harm due to false allegations (Milo, 2008). Historically, making damaging allegations against someone is costly to an accuser; the latter risks exposing themselves to threats, physical harm, lawsuits, and other potential social ramifications (e.g., Hershkowitz et al., 2021). The impersonal nature of ChatGPT challenges this condition; without embodiment and a way to ascribe "authorship" to the machine, ChatGPT's "accusations" could be more insidious given the low-cost nature for individuals or parties to generate and make false allegations.

The primary legal conundrum concerns one question: what is OpenAI's role in ChatGPT's misstep? Whether OpenAI should be held accountable in such cases involves legally determining whether the company is the *publisher* of defamatory material. Similar legal cases launched against Google and YouTube in Australia suggest that making such an argument involves a complex legal maneuver. For instance, in a case involving Google search results, the High Court of Australia did not regard the company as a publisher of the websites it links (High Court Judgments Database, 2022). Similarly, much of ChatGPT's training data are not produced by OpenAI, and the company could make a similar argument that ChatGPT is more like a distributor or "bookstore"—while it contains work that may be dangerous or false, it does not amount to being an "author." As the lawsuit was eventually abandoned (Swan, 2024), there remains an open question regarding the legal perception of responsibility regarding AI-generated misinformation.

In light of these ongoing cultural and legal debates, we must consider which groups in society have the legal resources or know-how to act on false information generated about them. While ChatGPT is more equipped to answer biographical questions about prominent people, making this population particularly sensitive to inaccurate outputs, they are also more likely to have the legal resources to seek justice for themselves. Since ChatGPT's release, dialogue agents have become increasingly linked to search—what happens when an employer uses ChatGPT to inquire about a prospective employee's personal background or history and fails to perform the due diligence of fact-checking? Those who belong to underresourced groups with limited access to corrective or legal measures may bear the brunt of ChatGPT's fabrications.

### A turn away from Turnitin? impact on higher ed teaching & learning

*Case study.* Ethan Mollick, an Associate Professor at the University of Pennsylvania's Wharton School, takes a radical approach to ChatGPT inclusion in the classroom (Kelly, 2023). Mollick's decision to include the dialogue agent as an assistive research tool in his syllabus just weeks after its launch contrasts with other educators' crackdowns on the use of the technology. Mollick's students had to work with ChatGPT to write, including learning to refine their prompts and bouncing ideas off the dialogue agent. Although Mollick admits to approaching ChatGPT with both enthusiasm and anxiety, he sees it as an unavoidable element of his students' futures in education. "The truth is, I probably couldn't have stopped them even if I didn't require it," Mollick said. A January 2023 survey by Study.com indicated that while 21% of educators had used ChatGPT to support their teaching in some capacity (by creating lesson plans, providing writing prompts, teaching writing styles, or operating as a digital tutor), more than 89% of students admitted to using ChatGPT to complete homework assignments and 48% had used it on a test, quiz, or essay (Study.com, 2023).

The rapid spread of ChatGPT use across educational systems necessitated development not only in classroom policy but also in technical support. On January 13, 2023, less than two months after ChatGPT was made publicly available, the plagiarism detection program Turnitin announced it was developing new tools to detect AI writing (Chechitelli, 2023). Calling such work "misconduct," it promised its own AI team was working hard to keep detection services at pace with generative models. A long-standing tool for educators to catch dishonest students, Turnitin was forced onto new terrain as educators and their students struggled to control the tools for writing.

### Societal impact analysis

Cheating and plagiarism in higher education is an industry of its own (Walker and Townley, 2012). Some students continuously find ways to get academic writing done, from hiring essay writers online to lifting paragraphs verbatim from internet sources. Unfortunately, despite the pervasive nature of the problem, identifying instances of plagiarism is also notoriously complex; instructors either rely on manual techniques (e.g., by comparing work to a student's earlier writing) or use a plagiarism detection service like Turnitin to check students' submissions against a database of previously submitted work and other digital sources.

What does ChatGPT mean for this uneasy truce? Almost immediately after its release, educators began sounding the alarm about ChatGPT's threat to student assessment. OpenAI did not initially release any accompanying services to detect text ChatGPT generates or rely on watermarking techniques, leaving educators scrambling to find a solution for a new kind of plagiarism. Before the emergence of dialogue agents like ChatGPT, educators generally trusted Turnitin to catch students who were not writing their own work, and warning students that Turnitin would evaluate their work deterred many would-be plagiarizers (Heckler et al., 2013). Turnitin's long-term usefulness may hinge on its ability to incorporate new techniques to evolve along with dialogue agents (e.g., to shift away from similarity checks to examine the "origin of content," as suggested by Khalil and Er, 2023).

Because OpenAI did not release plagiarism software alongside ChatGPT's release, AI-detection startups emerged to fill the gap. For example, GPTZero is a classification model launched in January 2023 that predicts whether a document was written with ChatGPT by comparing the variation and complexity of sentences (Ofgang, 2023). Although GPTZero was cited by some outlets as a relatively reliable AI detector (Wiggers, 2023), its creators cautioned against using the system to punish indicated plagiarism: "We recommend educators to take approaches that give students the opportunity to demonstrate their understanding in a controlled environment" (Tian, 2022). When OpenAI released its own detection tool two months after ChatGPT debuted, it came with similar words of caution: "Our classifier is not fully reliable….[it] correctly identifies 26% of AI-written text (true positives) as "likely AI-written," while incorrectly labeling human-written text as AI-written 9% of the time (false positives)" (OpenAI, 2023c).

Research on plagiarism detection in the ChatGPT era shows that the path forward remains uncertain. While Khalil and Er's (2023) study showed that only 20% of the essays generated by ChatGPT failed Ithenticate's (a Turnitin-like tool) plagiarism detection, Aydın and Karaarslan (2022) found the occurrence to be more frequent: 40%. Various factors, including the style, length, or topic of the essays evaluated, mediate the chances of Ithenticate labeling an original text as unoriginal. While the education sector awaits more sophisticated tools, and policies across institutions continue to vary, educators react in different manners to adapt how they evaluate their students' learning. For example, some advocate for more in-class writing assignments or oral assessments, while others design new projects that ChatGPT cannot handle (Rudolph et al., 2023).

Meanwhile, ad-hoc implementations of plagiarism detection strategies across classrooms in different locales may lead to unequal outcomes for various groups of students. Those more skilled at tinkering with ChatGPT's prompts and outputs could benefit more from using the technology than those less adept. The latest version of ChatGPT is only available to those paying for a premium subscription at USD 20 per month (OpenAI, 2023b). As usage increases, this might create an access divide among students. Moreover, the widespread use of imperfect detection services means that more students risk being accused of using ChatGPT even when they have not. Early research suggests students who are non-native English speakers are frequently misclassified by GPT detectors as plagiarizers (Liang et al., 2023). The deciding factors in whether or not the accusations stick will likely depend on a student's social and cultural capital (Strangfeld, 2019).

## To use or not to use: impact on workplace practices

*Case study.* ChatGPT is now a common workplace tool for workers across different industries. A February 2023 survey by FishBowl found that "70% of workers using ChatGPT at work [were] not telling" their employers (Graham, 2023). As a result, a new wave of data privacy issues has risen from workers inputting personal and sensitive data to ChatGPT. Some examples include doctors inserting patients' names and conditions into medical report prompts to enterprise workers using ChatGPT to draft business proposals containing proprietary information. Companies rushed to take action against such violations, with JP Morgan restricting its employees' use of ChatGPT, and Microsoft and Walmart advising caution and banning the sharing of "sensitive information" on such platforms (Lemos, 2023).

Many companies that allowed employee access to ChatGPT had to grapple with the consequences of such decisions. For example, while Amazon initially told workers that they could use ChatGPT if they were careful about sharing sensitive material, a company attorney later warned employees against sharing code with the dialogue agent after the enterprise reportedly witnessed ChatGPT responses reproducing internal Amazon data (Hurler, 2023). Similarly, after lifting an initial ban on ChatGPT, Samsung discovered that three engineers from the company's semiconductor division had input sensitive organizational information into ChatGPT (Dreibelbis, 2023). The incidents included an employee sharing source code from a semiconductor database, one attempting to identify defects in equipment by asking ChatGPT to diagnose its code, and another asking ChatGPT to generate minutes of an internal meeting.

### *Societal impact analysis*

ChatGPT's comprehensive set of capabilities has turned it into a one-stop shop for many work-related tasks (Chen et al., 2023). Today, workers use ChatGPT to write and summarize all sorts of documents, from thank-you emails to legal documents. The convenience of using one tool for multiple tasks makes ChatGPT particularly enticing, and its global popularity, alongside the widespread commentary on the potential of ChatGPT as a "copilot" (Philps and Tillman, 2023), places considerable pressure on employees to incorporate the tool into their workflows.

Such a workplace trend elevates information security risks for many organizations and corporations worldwide. For one, ChatGPT's free-to-use platform cultivates a reciprocal relationship with its users (Fourcade and Kluttz, 2020). Without monetary payment, what OpenAI gets in return is the copious amount of data individuals contribute to the system. Sharing sensitive and confidential information with publicly accessible systems like ChatGPT poses a significant risk to corporations like Samsung (Newman, 2023). OpenAI could use these sources of information to train subsequent systems and potentially appear in responses to other users in future iterations of the dialogue agent or other downstream applications. Ultimately, this risk led Samsung to outright ban employee access to ChatGPT (Gurman, 2023).

This worrying phenomenon may also disrupt protocols around data management. Facing disruptions to institutional norms around how proprietary data is used, employers must create policy on the fly as they determine, on the one hand, how to facilitate access to the tools that increase the productivity of their workforce and, on the other hand, how to retain sensitive information. Even companies that ban ChatGPT may face issues with compliance—employees found workarounds (such as using a VPN) to continue using ChatGPT to complete work-related tasks (e.g., The-Doodle-Dude, 2023). Evidently, preventing the use of ChatGPT in a climate where dependency has been cultivated is profoundly challenging.

In the early days after initial deployment, OpenAI could have offered a paid version of ChatGPT that did not retain any data users fed to the system. However, resource-constrained companies may not have sufficient funds to pay for access for every employee, essentially losing out on the ability to exploit the benefits of a new AI tool. Some more resourceful companies, like Samsung, may create their own private dialogue agents for employees (Gurman, 2023). In contrast, less resourceful entities may not have access to effective dialogue agents due to hardware, software, or other organizational barriers. As ChatGPT usage becomes more prevalent in the workplace, uneven access to privacy-preserving systems could reinforce inequalities between groups and organizations.

## Discussion

### *From human-centered to social-centered AI*

The potential for the recent proliferation of LLMs and dialogue agents to have long-standing impacts on society has been widely discussed by journalists and social scientists alike (Abdullah et al., 2022; Sanders and Schneier, 2023). We present the three disruptive events above to illustrate that existing pre- and post-deployment evaluation approaches for dialogue agents and their underlying LLMs fall short in identifying and responding to disruptions to groups and institutions, alongside their norms and practices. While some of these disruptions could lead to positive outcomes and further human–machine collaboration, the fruits they bear are generally unevenly distributed. Because the social cost of responding to ChatGPT's proliferation is placed on groups and institutions, those with more resources will be better equipped to tackle shifting needs.

Although some proponents of human-centered AI have suggested that the field of machine learning needs to consider humans at three "levels" or "spheres" (i.e., users, communities, and society; Landay, 2023), there is a lack of guidance on *how* to take social impact into account. Without clear directions, machine learning research communities predominantly draw on individualistic approaches stemming from human-centered principles to address the issues they are technically equipped to handle—for example, mitigating bias, toxicity, and "hallucination." These are important problems to solve, but the assumption that reducing individual-level harms translates directly to societal well-being results in poorly theorized conceptualizations of social impact. To address this gap, we propose a framework for social-centered AI that provides concrete recommendations for machine learning researchers and practitioners to consider the societal implications of their work.

### What does social-centered AI look like in practice?

In this section, we present a framework across different stages of AI development and flesh out the necessary organizational conditions to enable social-centered AI. Echoing Vaughan and Wallach's (2019) call for collaboration between machine learning researchers and human–computer interaction experts to realize a human-centered agenda, we stress the need for machine learning researchers and practitioners to engage with social scientists, who are more equipped to perform ethnographic and participatory research, to better understand the societal impacts of societal-scale AI systems. We target our interventions at large AI labs, often couched within big tech firms, as their privileged position makes them central to developing these large-scale systems.

### Phase 1: conceptualization & development

To counterbalance the tendency within machine learning research communities to prioritize technological solutionism (i.e., creating solutions in search of problems; Morozov, 2013), it is vital to collaborate with social scientists (e.g., sociologists, economists, anthropologists, etc.) during the early phases of project conceptualization to theorize and empirically examine the real-world problems AI systems could attenuate (Dinan et al., 2021; Ovadya and Whittlestone, 2019). Furthermore, for large-scale technologies that impact groups and institutions across various domains, it is vital to involve community stakeholders and identify key informants early to align goals through participatory research. The outcome of these collaborations between machine learning researchers, social scientists, and community stakeholders may generate questions that challenge existing agendas. For example, instead of multifunctional dialogue agents, would narrow AI—systems that are limited in scope, well-defined in their capabilities, and potentially less disruptive—better tackle the social problem at hand?

In our analysis, we show that from a technical and design perspective, displaying confidence scores on outputs, implementing watermarks and a plagiarism-detection application in synchrony with ChatGPT, and releasing a paid, privacy-preserving version of the dialogue agent early on could all be ways to better equip the dialogue agent for societal deployment. In addition, participatory research, such as focus groups with educators, could have elicited important insights and mitigation strategies to help OpenAI identify interest areas for a particular group of stakeholders. With such data in hand, machine learning researchers and practitioners can better direct engineering efforts and instigate measures that help society more adequately prepare for the deployment of high-impact technologies.

### Phase 2: short-term societal impact evaluation

Before a novel medical drug is approved for the general public, trials are performed on smaller groups of people to determine its safety and allow scientists to make any necessary changes to ensure responsible scaling. In such contexts, researchers are particularly sensitive to the needs of high-risk and vulnerable populations. We believe that societal-scale AI deployment should adopt a variation of this strategy—beta testing with specific populations that are likely to be heavily impacted rather than facilitating wide-scale access (e.g., Qadri et al., 2023). Indeed, before the release of ChatGPT, researchers from OpenAI noted the benefits of staggered releases of AI systems, which allow careful documentation and impact analysis that can be returned to developers for further iteration (Solaiman et al., 2019). We propose here that impact analysis during staggered releases goes beyond individual user feedback sourced from platform interfaces to produce descriptions of how people use a particular system in different real-world contexts.

An important limitation of this proposed strategy, however, is that researchers would retain the power to decide which people and communities to engage in staggered releases and on what terms—a common drawback of participatory approaches to computing and AI research (Cooper et al., 2022; Delgado et al., 2023). In addition, research suggests that some marginalized communities may be more amenable to techno-optimism, which may affect how those communities assess the risks and potential harms of sociotechnical systems (Kapania et al., 2022). We encourage further research to understand how to structure agreements between technology companies and partnering communities during this phase, to ensure the benefits of the work are shared, and to develop methodologies for engaging communities with a range of perspectives about the risks and harms of systems.

After wide release, societal-scale AI systems will likely impact various groups and institutions at disparate rates. While norms and practices are actively negotiated, AI researchers and practitioners must triage problems and prioritize focus areas. Here, relationships forged with community partners in the previous stage are vital. Developing infrastructure that allows the two sides to communicate easily would facilitate efficient feedback gathering from the most impacted communities and stakeholders. Platform features could be tweaked, altered, or added in the short term to allow for safer and more socially responsible adoption. It is important to acknowledge that predicting every possible short-term outcome is impossible, and there remains an ongoing challenge to translate localized, situation-specific analyses to scaled deployments of systems such as ChatGPT. However, prioritizing communicative channels that transcend individual-level feedback on a platform interface could help alleviate more immediate issues.

### Phase 3: longitudinal adaptations

Conducting social impact evaluations of LLMs and dialogue agents before and at the time of release will provide a cross-sectional view of shifts in norms and practice at a point in time. However, longitudinal studies are necessary for researchers to detect shifts or developments in groups and institutions over time, and to compare predictions before release with actual impacts at different time scales (Menard, 2002). Longitudinal studies also afford a more precise window into how different groups and institutions adapt to emerging technologies at different rates. Such studies are costly, requiring ongoing ethnographic analysis with affected people and quantitative analysis across different societal contexts. Nonetheless, insights from such analyses will be critical for shaping (or adapting) approaches to deploying future dialogue agents to maximize benefit for and minimize harm to affected groups and institutions (Shelby et al., 2023).

The collective feeling of the uncanny that many people experienced in the early days of ChatGPT's release, where fascination and fear intermingled (Gunning, 2008), was partly the result of a lack of shared knowledge or foresight on the future impact of the system on our social lives. To this end, implementing longitudinal studies investigating the impact of dialogue agents already deployed could prove instrumental. Such research would underscore the importance of social-centeredness within machine learning research communities, promoting it as a key research domain. Furthermore, this culture of social-centeredness will be essential to support broader initiatives to establish institutions and enforce regulations designed to oversee the deployment of future LLMs and dialogue agents.

### Organizational and regulatory considerations

We acknowledge that implementing the steps above will require changes in organizational and regulatory conditions and accordingly make the following recommendations. First, AI labs should prioritize an interdisciplinary work environment; while technical workers such as research scientists and engineers are well-equipped to handle feedback provided by individual users, social evaluations on a larger scale demand investments in those with different skill sets (Kusters et al., 2020; Selbst et al., 2019). Engendering a collective sense of "algorithmic realism" (Green and Viljoen, 2020) necessitates true interdisciplinarity, but this interdisciplinarity faces a significant obstacle in the unequal power distribution within large AI labs. As Abbott (2014) notes, professional groups often control expert knowledge creation through hierarchical power structures. Meaningful interdisciplinary collaboration requires that labs empower social scientists with real, actionable power in decision-making across development and deployment phases, including planning, data collection, modeling, and evaluation.

Second, realizing social-centered AI requires incentive alignment. Current priorities such as profit motives, rapid deployment to edge out competition, and protecting intellectual property are all factors that could impede the implementation of social-centered AI. However, building on recent initiatives such as the NeurIPS broader impact statements (Ashurst et al., 2022) to compel researchers to reflect more comprehensively and critically about social impacts in professional conferences and journal submissions will be an important step toward a culture where social-centered approaches become the norm rather than the exception.

The third and perhaps most important dimension of realizing social-centered AI is governance (Dafoe, 2018). More specifically, we need to build institutions and governmental agencies that create policies and regulatory standards—mechanisms that structurally reinforce behavior. Recent initiatives show that governments are increasingly recognizing the societal impacts of AI and the inadequacy of existing legal and policy arrangements to deal with those impacts. For example, the European Union AI Act includes strict obligations for reporting, evaluating, and ongoing assessment of "high-impact general purpose AI" models with systemic risk (European Parliament, 2024). With the realization of the steps outlined earlier, the data gathered by social scientists (in partnership with community stakeholders) will be invaluable for AI labs to meet these obligations and for policymakers to understand and respond to the social impacts of such systems (Rakova and Dobbe, 2023).

## Conclusion

In this paper, we propose complementing the dominant, individualistic framing of human-centered AI used by machine learning research communities with a social-centered AI paradigm. We underscore the urgency of this shift by presenting three case studies of disruptive events engendered by ChatGPT, which illustrate the discrepancy between technical evaluations and public concerns regarding changes to the norms and practices of social groups and institutions. These case studies highlight the uneven distribution of benefits of dialogue agents and their underlying LLMs, determined by the relative access to resources of users or stakeholders. To better understand and address these impacts of societal-scale AI systems, we advocate for integrating ethnographic analyses and participatory approaches in future AI development and deployment.

As future iterations of AI systems include an increasing number of modalities and languages, their impact on society will be more widely felt. In an op-ed titled "This Changes Everything" in March 2023, columnist Klein (2023) argued that "One of two things must happen. Humanity needs to accelerate its adaptation to these technologies or a collective, enforceable decision must be made to slow the development of these technologies." By proposing social-centered AI, we chart a path toward the latter objective (see Peng and Zhao 2024). We acknowledge that building a culture around the concept will incontrovertibly slow down the developmental progress of AI, but we believe that it is a necessary trade-off to ensure the responsible and ethical integration of its systems into society over time.

## Positionality

The authors of this article are interdisciplinary researchers working on topics related to the intersection of technology and society. While we have diverse cultural and professional backgrounds, we are all presently nested in high-intensity research institutions in the Global North. Two authors also have experience in the research divisions of large technology companies, focusing on Responsible AI. Our positions afford us access to resources and networks that few users or stakeholders of dialogue agents and LLMs experience.

While we approach our work with a critical lens, we acknowledge our position within the academic-industrial complex. Our perspective is shaped by a commitment to improving the deployment process rather than advocating for the abolition of AI systems. This stance stems from a belief in the potential benefits of such systems for many users, combined with recognizing the importance of social-centered development practices. Others occupying different positions—for example, those outside the academic-industrial complex affected negatively by recent deployments—might have a more critical stance toward such artifacts.

## Acknowledgments

The authors are indebted to Eun Seo Jo for her early contributions to this work, and to Anastasia Nikoulina, Rida Qadri, Mary Gray, and Angèle Christin for the generative discussions that shaped various aspects of the paper. The authors also thank Luis Tenorio, Glen Berman, David Joseph-Goteiner, Nataliya Nedzhvetskaya, and Alex Zafiroglu for their valuable feedback.

## Declaration of conflicting interests

The author(s) declared no potential conflicts of interest with respect to the research, authorship, and/or publication of this article.

## Funding

The author(s) disclosed receipt of the following financial support for the research, authorship, and/or publication of this article: Ned Cooper is supported by an Australian Government Research Training Program (RTP) Scholarship and a Florence Violet McKenzie scholarship.

## ORCID iDs

Skyler Wang https://orcid.org/0000-0002-0639-945X
Ned Cooper https://orcid.org/0000-0003-1834-279X
Margaret Eby https://orcid.org/0000-0003-1004-5934

## Note

1. We use the term "hallucinate" in this paper, reflecting its wide use in natural language generation literature (Ji et al., 2023), though we recognize "confabulate" is a more accurate term for the phenomenon (Smith et al., 2023).